\documentclass{aa6}

\usepackage{graphicx}
\usepackage{natbib}

\newcommand{\id}{\mbox{$\mathrm{d^{-1}}$}}
\newcommand{\Msun}{\mbox{$\mathrm{M}_{\odot}$}}
\newcommand{\Rsun}{\mbox{$\mathrm{R}_{\odot}$}}
\newcommand{\kms}{\mbox{$\mathrm{km\,s^{-1}}$}}
\newcommand{\Mwd}{\mbox{$M_\mathrm{wd}$}}
\newcommand{\Msec}{\mbox{$M_\mathrm{sec}$}}
\newcommand{\Mone}{\mbox{$M_\mathrm{1}$}}
\newcommand{\Rwd}{\mbox{$R_\mathrm{wd}$}}
\newcommand{\Rsec}{\mbox{$R_\mathrm{sec}$}}

\newcommand{\Kwd}{\mbox{$K_\mathrm{wd}$}}
\newcommand{\Kseccor}{\mbox{$K_\mathrm{sec,cor}$}}
\newcommand{\Kseccal}{\mbox{$K_\mathrm{sec,cal}$}}
\newcommand{\Twd}{\mbox{$T_\mathrm{wd}$}}
\newcommand{\Tsec}{\mbox{$T_\mathrm{sec}$}}
\newcommand{\Tone}{\mbox{$T_\mathrm{1}$}}

\newcommand{\Kem}{\mbox{$K_\mathrm{em}$}}

\newcommand{\T}{\mbox{$T_\mathrm{0}$}}

\newcommand{\Porb}{\mbox{$P_{\rm orb}$}}
\newcommand{\Line}[3]{\Ion{#1}{#2}\,\,$\lambda$#3}

\newcommand{\Idline}[2]{\Ion{#1}{#2}}
\newcommand{\Ion}[2]{#1{\,\scriptsize #2}}
\newcommand{\Ha}{\mbox{${\mathrm H\alpha}$}}
\newcommand{\Hb}{\mbox{${\mathrm H\beta}$}}
\newcommand{\Hg}{\mbox{${\mathrm H\gamma}$}}
\newcommand{\Hd}{\mbox{${\mathrm H\delta}$}}
\newcommand{\He}{\mbox{${\mathrm H\epsilon}$}}
\newcommand{\Hten}{\mbox{${\mathrm H10}$}}
\newcommand{\Hthirteen}{\mbox{${\mathrm H13}$}}

\begin{document}

\title{HS\,1857+5144: A hot and young pre-cataclysmic variable}

\author{A. Aungwerojwit\inst{1,2}\and 
        B. T. G\"ansicke\inst{1}\and
        P. Rodr\'iguez-Gil\inst{3}\and
        H.-J. Hagen\inst{4}\and
	O. Giannakis\inst{5}\and
	C. Papadimitriou\inst{5}\and
	C. Allende Prieto\inst{6}\and
        D. Engels\inst{4}
        }
\authorrunning{Aungwerojwit et al.}
\titlerunning{HS\,1857+5144: A hot and young pre-cataclysmic variable}

\offprints{A. Aungwerojwit, \\ e-mail: A.Aungwerojwit@warwick.ac.uk}

\institute{
   Department of Physics, University of Warwick, Coventry, CV4 7AL, UK
\and
   Department of Physics, Faculty of Science, Naresuan University, 
   Phitsanulok, 65000, Thailand
\and
   Instituto de Astrof\'isica de Canarias, 38200 La Laguna, Tenerife, Spain
\and
   Hamburger Sternwarte, Universit\"at Hamburg, Gojenbergsweg
   112, 21029 Hamburg, Germany
\and 
   Institute of Astronomy and Astrophysics,
   National Observatory of Athens, P.O. Box 20048, Athens 11810, Greece
\and
McDonald Observatory and Department of Astronomy, University of
   Texas, Austin, TX 78712, USA
}

\date{Received \underline{\hskip2cm} ; accepted }

\abstract{}
{We report the discovery of a new white dwarf/M dwarf binary,
HS\,1857+5144, identified in the Hamburg Quasar Survey (HQS).}
{Time-resolved optical spectroscopy and photometry were carried out to
determine the properties of this new cataclysmic variable progenitor
(pre-CV).}
{The light curves of HS\,1857+5144 display a sinusoidal variation with
a period of $\Porb=383.52$\,min and peak-to-peak amplitudes of 0.7\,mag and
1.1\,mag in the $B$-band and $R$-band, respectively. The large
amplitude of the brightness variation results from a reflection effect
on the heated inner hemisphere of the companion star, suggesting a
very high temperature of the white dwarf. Our radial velocity study
confirms the photometric period as the orbital period of the system. A
model atmosphere fit to the spectrum of the white dwarf obtained at minimum
light provides limits to its mass and temperature of
$\Mwd\simeq0.6-1.0$\,\Msun\ and $\Twd\simeq70\,000-100\,000$\,K,
respectively. The detection of \Line{He}{II}{4686} absorption
classifies the primary star of HS\,1857+5144 as a DAO white
dwarf. Combining the results from our spectroscopy and photometry, we
estimate the mass of the companion star and the binary inclination to
be $\Msec\simeq0.15-0.30$\,\Msun\ and $i\simeq45\degr-55\degr$,
respectively.}
{We classify HS\,1857+5144 as one of the youngest pre-CV known to
date. The cooling age of the white dwarf suggests that the present
system has just emerged from a common envelope phase $\sim10^{5}$\,yr
ago. HS\,1857+5144 will start mass transfer within or below the
2--3\,h period gap.}
\keywords{stars: binaries: close -- stars: individual: HS\,1857+5144-- stars:
  pre-cataclysmic variables}

\maketitle

\section{Introduction}
Post-common envelope binaries (PCEBs), i.e. detached white dwarf-main
sequence binaries, originate from wide binaries comprising unequal
main sequence components. Once the more massive star evolves through
the giant phase and fills its Roche lobe, unstable mass transfer onto
the unevolved star starts. The high accretion rate subsequently brings
the system into a common envelope phase (CE). Friction between the
stellar components and the envelope shrinks the binary orbital
separation and eventually ejects the envelope from the system,
resulting in the majority of PCEBs having orbital periods of a few
days \citep{willems+kolb04-1}. The binary separations of PCEBs are
believed to be further reduced through angular momentum loss via
magnetic braking and/or gravitational radiation.  The main-sequence
star will ultimately fill its Roche lobe and start mass transfer onto
the white dwarf, which turns the binary into a cataclysmic variable
(CV). While this general scenario is widely accepted, the details of
the evolution through the common envelope phase, as well as the
subsequent orbital angular momentum loss are poorly
understood. Consequently predictions made by binary population models
are rather uncertain.

As PCEBs are simple objects in terms of their stellar components, they
offer a large potential in constraining and calibrating the physics of
both CE evolution and orbital angular momentum loss. This has
implications for a wide range of astronomical objects such as SN\,Ia
progenitors, X-ray binaries, or neutron star binaries as likely
progenitors of short gamma-ray bursts. However, until recently PCEBs
received little observational attention, largely due to the lack of a
dedicated search for such systems. \citet{schreiber+gaensicke03-1}
analysed the properties of 30 well-studied PCEBs and showed that the
known population of these systems is extremely biased towards
young systems consisting of hot white dwarfs and late type companions, as
the majority of the known PCEBs were initially selected as blue
objects.  A substantial improvement in the statistics of PCEB
properties will be possible through the exploitation of large
extragalactic surveys such as the Hamburg Quasar Survey (HQS,
\citealt{hagenetal95-1}) or the Sloan Digital Sky Survey (SDSS,
\citealt{yorketal00-1}).

Here we present a detailed follow-up study of the new PCEB
HS\,1857+5144, which has been discovered in our ongoing effort to
identify CVs and pre-CVs in the HQS (see
\citealt{gaensickeetal02-2, aungwerojwitetal05-1} for details
on the project). In Sect.\,\ref{s-observations} we describe the
observations and data reduction. The orbital ephemeris of
HS\,1857+5144 is determined in Sect.\,\ref{s-analysis}. In
Sects.\,\ref{s-stellar_components}--\ref{s-discussion} we analyse the
nature of the stellar components and discuss the future evolution of
the system.

\begin{figure}
\centerline{\includegraphics[width=6.8cm]{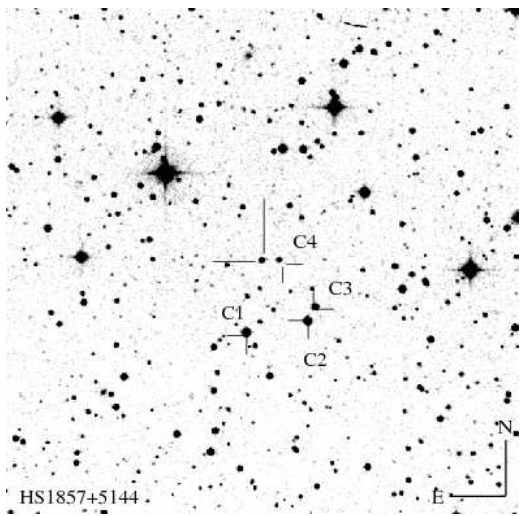}}
\caption{\label{f-fc} A $10\arcmin\times10\arcmin$ finding chart for
HS\,1857+5144 obtained from the Digitized Sky Survey. The J2000
coordinates of the star are
$\alpha=18^\mathrm{h}58^\mathrm{m}32.12^\mathrm{s}$ and
$\delta=+51\degr48\arcmin57.74\arcsec$. The comparison and check
stars used in the photometry are marked by `C1'-`C4' (see
Sect.\,\ref{s-observations_phot} for details).}
\end{figure}

\section{Observations and Data Reduction\label{s-observations}}
\subsection{Spectroscopy\label{s-obsspect}}
An identification spectrum of HS\,1857+5144 was obtained in August
1992 with the Boller \& Chivens Cassegrain spectrograph on the 2.2-m
telescope at Calar Alto Observatory. The spectrum is characterised by
a blue continuum superimposed by strong Balmer emission
lines. Subsequent time-series, intermediate resolution spectroscopy of
HS\,1857+5144 was performed in July 2004 at the 2.7-m Harlan J. Smith
telescope at McDonald Observatory equipped with the Large Cassegrain
Spectrograph (LCS), covering $\sim10$\,h (35 spectra) in total. The
individual spectra were obtained through a 1\arcsec slit and grating
\#43 and imaged on the $800 \times 800$ pixel TI1 CCD camera. This
setup provided access to the $\lambda\lambda3670-5050$ wavelength
range at 3.5 \AA~spectral resolution. The reduction and optimal
extraction of the spectra were performed using standard long-slit
spectroscopy packages within \texttt{IRAF}\footnote{\texttt{IRAF} is
distributed by the National Optical Astronomy Observatories, which are
operated by the Association of Universities for Research in Astronomy,
Inc., under cooperative agreement with the National Science
Foundation.}. Four additional high resolution spectra were obtained
using the Intermediate Dispersion Spectrograph and Imaging System
(ISIS) on the 4.2-m William Herschel Telescope (WHT) on La Palma in
July 2006, covering the orbital minimum, maximum, and the quadrature
phases. The blue arm of the spectrograph was equipped with the R1200B
grating, providing a spectral resolution of $\simeq1$\,\AA\ over the
wavelength range $4200-5000$\,\AA. The data were reduced using the
\texttt{Figaro} within the 
\texttt{Starlink} package as well as \texttt{Pamela} and
\texttt{Molly} written by
T. Marsh\footnote{www.warwick.ac.uk/go/trmarsh}.

\begin{figure*}
\centerline{\includegraphics[angle=-90,width=15cm]{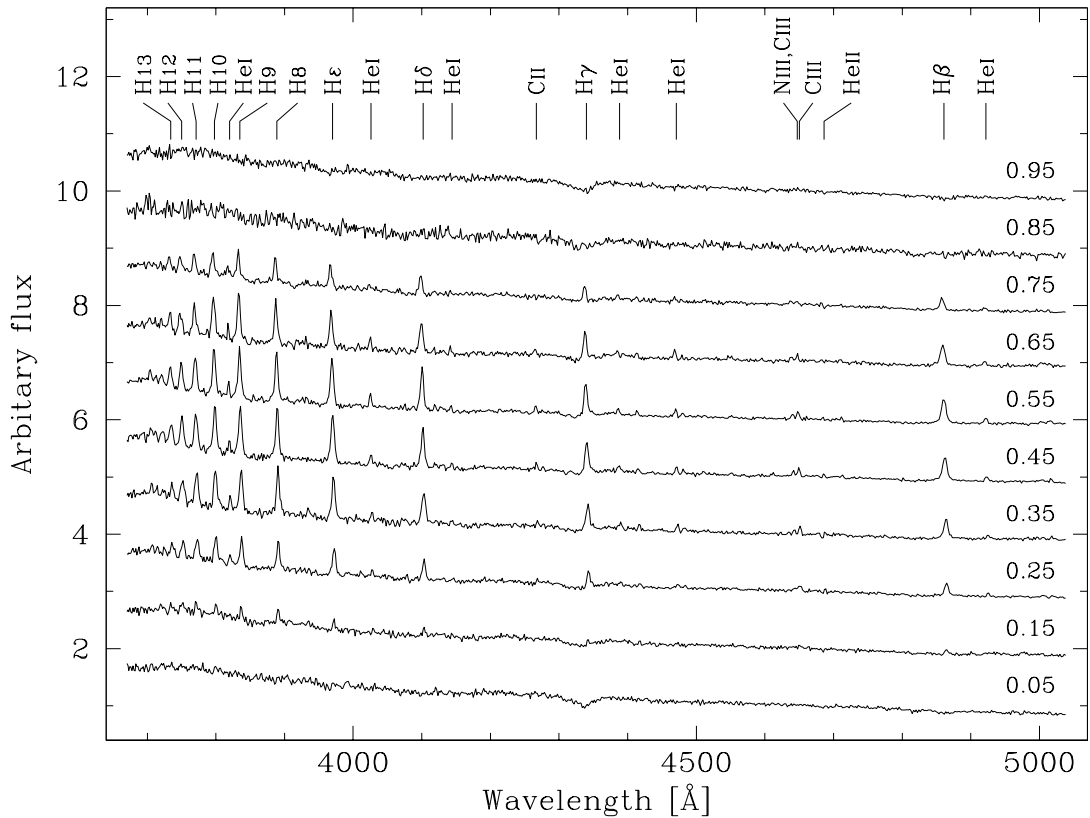}}
\caption{\label{f-phasebinmc} Phase-binned spectra of HS\,1857+5144 obtained at
McDonald Observatory. These spectra show a clear modulation of the emission
line strengths with orbital phase.}
\end{figure*}

\begin{figure*}
\centerline{\includegraphics[angle=-90,width=15cm]{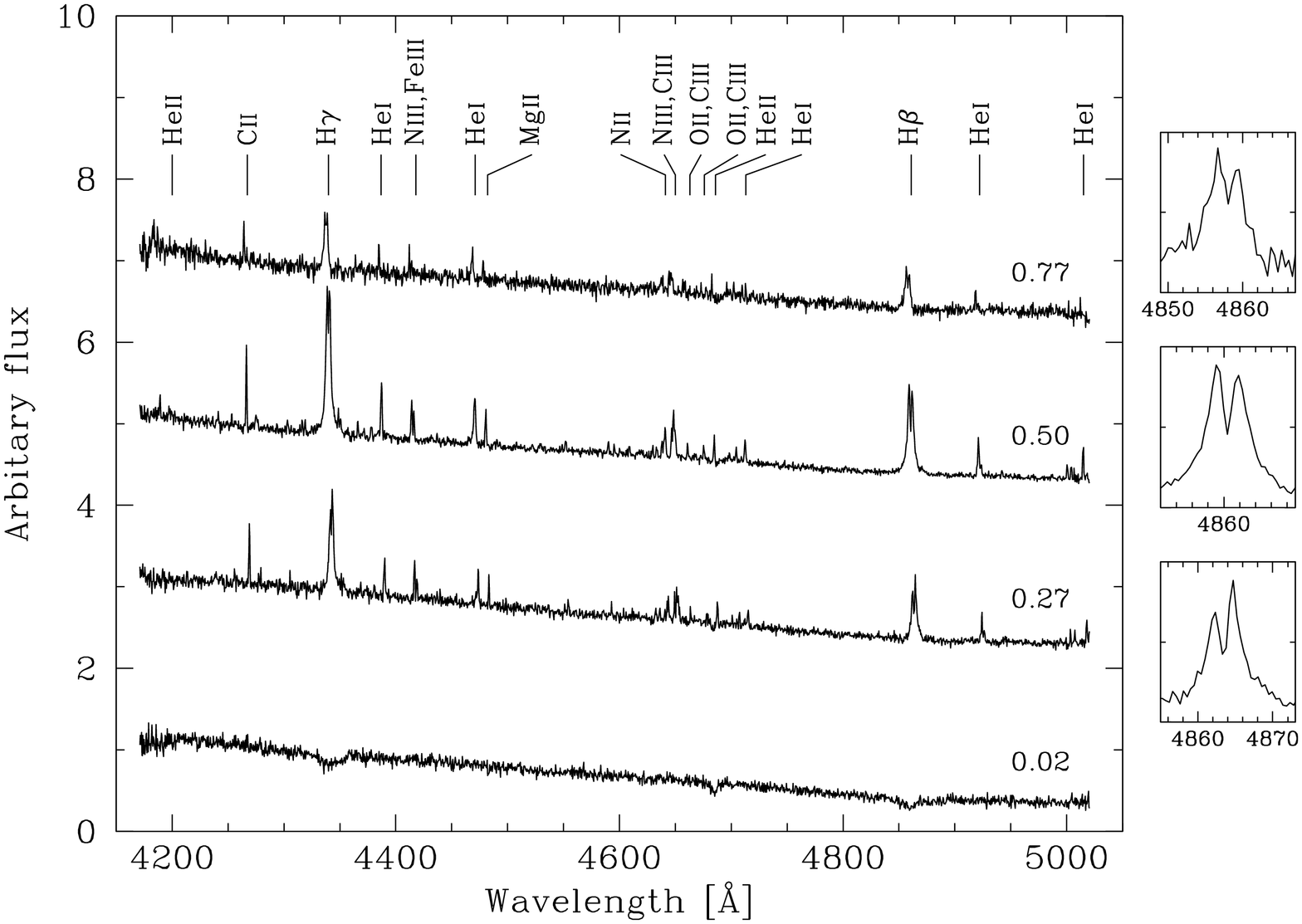}}
\caption{\label{f-phasebinwht} {\it Main panel}: high resolution spectra of
HS\,1857+5144 at different orbital phases from $0.02-0.77$ obtained at
the WHT. {\it Small windows}: close up of the evolution of the \Hb\
emission line profile in the orbital phase range $0.27-0.77$ (from
{\it bottom} to {\it top}).}
\end{figure*}

\paragraph{Spectroscopic characteristics.}
Figures\,\ref{f-phasebinmc} and \ref{f-phasebinwht} illustrate the
orbital phase-dependent variation of the emission line profiles of
HS\,1857+5144 from the McDonald and WHT spectra, respectively.  The
emission lines greatly vary in strength over the orbital cycle, with
maximum line fluxes occurring at $\varphi\simeq0.5$ and disappearing
around $\varphi\simeq0.0$. The dominant lines detected in the McDonald
spectra belong to the Balmer series, but the high quality of the WHT
spectra reveals a multitude of narrow emission lines, e.g. the
\Idline{C}{III}/\Line{N}{III}{4650}\ Bowen blend, \Idline{Mg}{II},
\Idline{N}{II}, \Idline{O}{II}/\Idline{C}{III}\ blend, and
\Idline{N}{III}/\Idline{Fe}{III}\ blend. This emission spectrum
is a characteristic of known PCEBs containing a cool secondary star
irradiated by a very hot primary component, such as BE\,UMa
\citep{fergusonetal81-1, ferguson+james94-1}, EC\,11575-1845
\citep{chenetal95-1}, and HS\,1136+6644 \citep{singetal04-1}.  The
strengths of the Balmer emission lines vary in phase with those of
\Idline{He}{I}, \Idline{C}{III}/\Idline{N}{III}, indicating that all
emission lines come from the same source.  The WHT spectra also reveal
that the Balmer emission lines have double-peaked profiles, with peak
separations of $\sim170$\,\kms\ for \Hb\ (Fig.\,\ref{f-phasebinwht},
small window) and $\sim150$\,\kms\ for \Hg\, which are most likely
caused by non-LTE effects in the strongly irradiated atmosphere of the
companion star \citep{barmanetal04-1}.

During the orbital faint phase, $\varphi\simeq0.0$, weak absorption
Balmer lines are observed in both sets of spectra, with
\Line{He}{II}{4686} absorption also detected in the WHT
spectra. The detection of \Line{He}{II}{4686} classifies the primary star
in HS\,1857+5144 as a DAO white dwarf.

\subsection{Photometry}
\label{s-observations_phot}
\begin{figure}
\centerline{\includegraphics[angle=-90,width=\columnwidth]{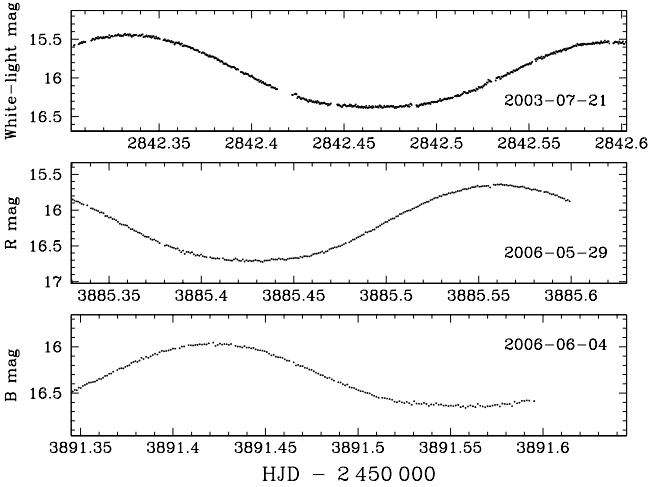}}
\caption{\label{f-lc_hs1857} Filterless, $R$, and $B$ (from {\it top}
to {\it bottom}) sample light curves of HS\,1857+5144 obtained with
the 1.2-m telescope at Kryoneri Observatory.}
\end{figure}

We obtained a total of $\sim60$\,h of time-series CCD photometry of
HS\,1857+5144 (Fig.\,\ref{f-fc}) during the period July 2003 to June
2006 (Table\,\ref{t-obslog}). Filterless photometry was carried out in
2003/4 using the 1-m Optical Ground Station (OGS) and the 0.82-m IAC80
telescope at the Observatorio del Teide on Tenerife. Both telescopes
were equipped with Thomson $1\mathrm{k}\times1\mathrm{k}$ pixel CCD
cameras. Photometric observations were also carried out with the 1.2-m
telescope at Kryoneri Observatory and a $516\times516$ pixel
Photometrics SI-502 CCD camera. The OGS data were reduced in a
standard fashion with \texttt{IRAF}, and the instrumental magnitudes
of the object and comparison stars in the field were extracted using
the point spread function (PSF) packages. Differential magnitudes of
HS\,1857+5144 were then computed relative to the comparison star `C1'
(USNO-A2.0\,1350-10080469: $R=13.2$, $B=14.7$) whose brightness
variation was found to be negligible against the check star `C2'
(USNO-A2.0\,1350-10078502: $R=13.5$, $B=14.6$).  The IAC80 and
Kryoneri data were reduced using the pipeline described by
\citet{gaensickeetal04-1} which employs \texttt{MIDAS} for bias and
dark current subtraction and flat fielding, and performs aperture
photometry over all visible objects using \texttt{Sextractor}
\citep{bertin+arnouts96-1}. Differential magnitudes for HS\,1857+5144
were calculated from the Kryoneri data using again the comparison star
`C1'and the check star `C2'. For the IAC80 data, the comparison star
`C3' (USNO-A2.0\,1350-10078272: $R=14.5$, $B=16.1$) and the check star
`C4' (USNO-A2.0\,1350-10079362: $R=16.4$, $B=17.8$) were used.
Additional $B$- and $R$-band light curves of HS\,1857+5144 were
obtained at Kryoneri observatory in May/June 2006 and reduced in the
same way as the filterless data from this telescope. Sample of white
light, $B$- and $R$-band light curves are displayed in
Fig.\,\ref{f-lc_hs1857}.

\paragraph{Light curve morphology.}
The light curves of HS\,1857+5144 (Fig.\,\ref{f-lc_hs1857}) display a
smooth quasi-sinusoidal modulation with a period of $\simeq6.4$\,h and
peak-to-peak amplitudes of 0.7\,mag in the $B$-band, 1.1\,mag in the
$R$-band, and $0.9$\,mag in white light. The minimum in the $B$-band
light curve is nearly flat for $\simeq0.15$ orbital cycle, whereas the
shape of the minimum in the $R$-band is rounder. No sign of the
typical short-period flickering of accreting systems is detected,
which classifies HS\,1857+5144 as a detached binary. The low-amplitude
scatter seen in the light curves in Fig.\,\ref{f-lc_hs1857} is caused
by residual flat field structures and poor tracking of the Kryoneri
telescope. The observed periodic brightness variation is
a characteristic of a large reflection effect on the heated face of the
secondary star, irradiated by a hot primary star (e.g. TW\,Crv,
\citealt{chenetal95-1}; KV\,Vel, \citealt{hilditchetal96-1}; and
HS\,2333+3927, \citealt{heberetal04-1}).  Finally, HS\,1857+5144 was
found at a constant mean magnitude of $\simeq16.2$ throughout our
observing runs, consistent with USNO-A2.0 measurements of
HS\,1857+5144 ($R=16.3$ and $B=15.7$).

\begin{table}[t]
\caption[]{Log of the observations\label{t-obslog}.}
\setlength{\tabcolsep}{0.95ex}
\begin{flushleft}
\begin{tabular}{lccccc}
\hline\noalign{\smallskip}
Date & UT &  Telescope & Filter/ & Exp. & Frames  \\
     &    &            & Grism   & (s)  &          \\
\hline\noalign{\smallskip}
1992 Aug 10 & 21:51       & CA2.2 &       &  1500  & 1 \\
2003 Jul 10 & 21:03-01:01 & OGS   & clear &  17    & 610 \\
2003 Jul 13 & 21:16-01:00 & OGS   & clear &  12    & 751 \\
2003 Jul 21 & 19:15-02:25 & KY    & clear &  30    & 671 \\
2003 Jul 23 & 18:56-02:32 & KY    & clear & 20-45  & 311 \\
2004 May 21 & 01:43-05:25 & IAC80 & clear &  40    & 231 \\
2004 May 23 & 00:52-02:07 & KY    & clear & 20-30  & 199 \\
2004 May 25 & 21:34-02:26 & KY    & clear &  20    & 587 \\
2004 May 26 & 23:14-02:29 & KY    & clear &  20    & 335 \\
2004 May 27 & 02:51-05:12 & IAC80 & clear &  40    & 286 \\ 
2004 Jun 09 & 21:40-02:20 & KY    & clear &  20    & 608 \\
2004 Jul 16 & 04:13-11:01 & McD   & \#43  & 600    &  31  \\            
2004 Jul 19 & 04:25-08:58 & McD	  & \#43  & 600    &   4 \\
2006 May 28 & 21:21-02:14 & KY    & $R$   & 70-80  & 212 \\
2006 May 29 & 19:53-02:21 & KY    & $R$   &   80   & 267 \\
2006 May 30 & 20:58-02:21 & KY    & $R$   &   80   & 186 \\
2006 Jun 04 & 20:15-02:14 & KY    & $B$   & 90-110 & 196 \\
2006 Jul 02 & 21:17       & WHT   & R1200 & 600    & 1   \\
2006 Jul 03 & 21:21       & WHT   & R1200 & 600    & 1   \\
2006 Jul 04 & 21:17       & WHT   & R1200 & 600    & 1   \\
2006 Jul 05 & 21:16       & WHT   & R1200 & 600    & 1   \\
\noalign{\smallskip}\hline
\end{tabular}
\end{flushleft}
Notes. CA2.2m: 2.2-m telescope, Calar Alto Observatory; IAC80:
0.82-m telescope, Observatorio del Teide; KY: 1.2-m
telescope, Kryoneri Observatory; McD: 2.7-m Harlan J. Smith
Telescope, McDonald Observatory; OGS: 1-m Optical Ground Station,
Observatorio del Teide; WHT: 4.2-m William Herschel
Telescope, Roque de Los Muchachos Observatory.
\end{table}

\section{Analysis}\label{s-analysis}
\subsection{Orbital period and ephemeris}\label{s-porb}
\begin{figure}
\centerline{\includegraphics[angle=-90,width=\columnwidth]{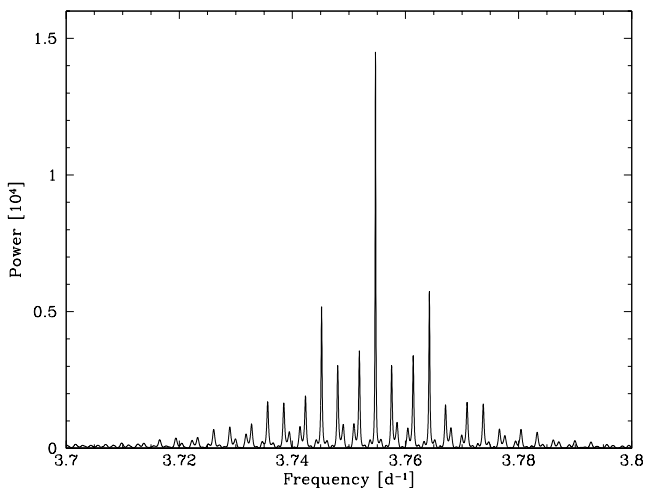}}
\caption{\label{f-scargle_hs1857}The \texttt{ORT} periodogram of
HS\,1857+5144 computed from all photometric data.}
\end{figure}

A qualitative inspection of the light curves presented in
Sect.\,\ref{s-observations_phot} provided an estimate of the orbital
period of $\simeq6.4$\,h. In order to obtain a precise value, we
shifted the magnitudes of each observing run so that their minima
match that of the 2003 July 21 Kryoneri data, which covered an entire
orbital cycle. We then subjected the combined 2003--2006 data to a
time-series analysis using Schwarzenberg-Czerny's
(\citeyear{schwarzenberg-czerny96-1}) \texttt{ORT} method, a variation
of the analysis-of-variance technique which fits orthogonal
polynomials to the data folded over a set of trial periods. The
\texttt{ORT} periodogram (Fig.\,\ref{f-scargle_hs1857}) contains an
unambiguous peak at 3.755\,\id. A sine fit to the combined photometric
data defined the following ephemeris:
\begin{equation}
\label{e-ephemeris}
\T=\mathrm{HJD}\,2452831.5475(17)+ 0.26633357(8)\times E~~,
\end{equation}
where \T\ is defined as the time of inferior conjunction of the
secondary star (=\,orbital minimum in the light curves). We conclude
that the orbital period of HS\,1857+5144 is
$\Porb=383.5203\pm0.0001$\,min. Figure\,\ref{f-fold_hs1857} (bottom two
panels) shows the Kryoneri $B$- and $R$-band light curves folded
according to the above ephemeris.

\subsection{Radial velocities and equivalent widths}\label{s-rv}
In order to spectroscopically confirm the orbital period of
HS\,1857+5144, we first measured radial velocity variations of the
\Hb, \Hg, \Hd, \He, and \Hten\ emission lines separately by
cross-correlating the observed line profiles with a single Gaussian
with a full-width at half-maximum (FWHM) of 250\,\kms\ for \Hb\ and
\He, and of 300\,\kms\ for \Hg, \Hd, and \Hten\ within
\texttt{MOLLY}. The \texttt{ORT} periodograms calculated from the radial
velocities of the individual lines consistently favoured an orbital
frequency of $\sim3.75$\,\id, in good agreement with the photometric
result. The radial velocity amplitudes determined from the different
Balmer lines varied in the range $\sim175-215\,\kms$.

In order to obtain a more robust measure of the radial velocity of the
companion star in HS\,1857+5144, we determined an average radial
velocity of the Balmer lines by fitting simultaneously the Balmer
series \Hb\ to \Hthirteen\ with a set of 10 Gaussians. The wavelengths
of all Gaussians were fixed to their laboratory wavelengths relative
to that of \Hd, and only the wavelength of \Hd, as well as the widths
and amplitudes of all 10 Gaussians were used as a free parameters. The
average Balmer line radial velocities are listed in
Table\,\ref{t-rv+ew} and are shown in Fig.\,\ref{f-fold_hs1857} (top
panel) folded over the ephemeris given in Eq.\,\ref{e-ephemeris}.  A
sine fit to the folded velocities {and their errors results in
an amplitude of $\Kem=185.2\pm4.9$\,\kms\ and
$\gamma=-24.0\pm6.5$\,\kms. Note that this velocity amplitude does not
represent the radial velocity amplitude of the centre of mass of the
secondary star, but that of the illuminated hemisphere. Since the
centre of light is located closer to the centre of mass of the system
than the centre of mass of the secondary star, the 'true' radial
velocity amplitude of the secondary star should therefore be larger
than the observed velocity amplitude, $\Kem=185.2\pm4.9$\,\kms. We
will determine a $K$-correction in Sect.\,\ref{s-kcorrection}.

We also analysed the variation of the equivalent width (EW) of the
\Hb\ line (Table\,\ref{t-rv+ew}). The Scargle periodogram
calculated from these measurements contained two equally significant
signals at $3.409$\,\id\ and $3.749$\,\id, the latter of which agrees
well with the orbital frequency derived from the photometry and from
the radial velocity variations. The equivalent width measurements
folded over Eq.\,(\ref{e-ephemeris}) are shown in
Fig.\,\ref{f-fold_hs1857} (second panel from top). As expected
for an irradiation effect, maximum equivalent width takes place at
$\varphi\simeq0.5$.

\begin{figure}
\centerline{\includegraphics[width=\columnwidth]{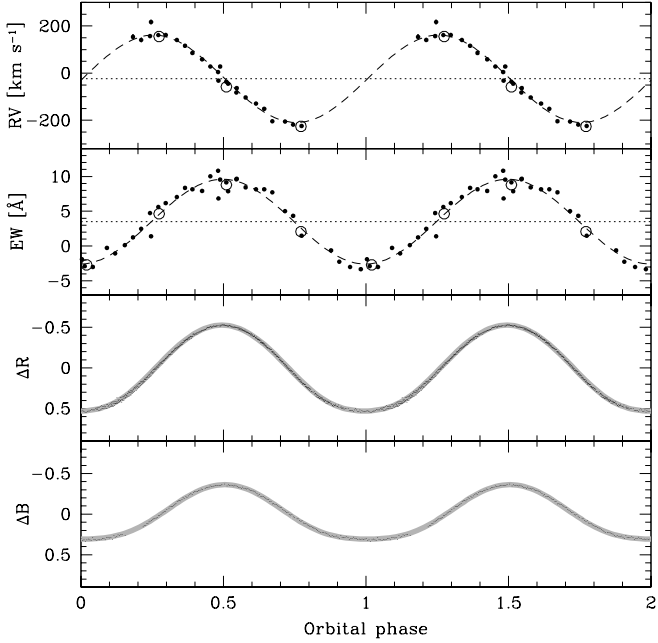}}
\caption{\label{f-fold_hs1857} Spectroscopic and photometric data of
HS\,1857+5144 folded over the photometric orbital period of 383\,min
given in Eq.\,\ref{e-ephemeris}. {\it Top two panels}: the average of
the Balmer radial velocities and \Hb\ equivalent width variations
along with the best sine fit (dashed line); the filled and open
circles represent the Mcdonald and WHT data, respectively. The
error bars in the radial velocity measurements are included in the
plot, but of similar size as the points. The uncertainties in the
values of the equivalent widths are dominated by systematic effects of
the order $\sim1$\,\AA. {\it Bottom two panels}: $R$-band and
$B$-band light curves obtained during May/June 2006 along with the
synthetic light curves (gray line) computed with the program
\texttt{PHOEBE} for $\Mwd=0.72$\,\Msun, $\Msec=0.21$\,\Msun,
$i=53\degr$, and $\Twd=100\,000$\,K. Phase zero is defined as inferior
conjunction of the secondary star. Note that phase of the radial
velocity curve is offset with respected to equivalent width variation
curve and light curves by $\simeq0.25$, consistent with an origin of
the emission lines on the heated inner hemisphere of the secondary
star (see Sect.\,\ref{s-phaserelations} for details).}
\end{figure}

\begin{table}[t]
\caption[]{The average radial velocities of the Balmer emission lines and
\Hb\ equivalent widths of HS\,1857+5144 measured from the McDonald and WHT
spectra \label{t-rv+ew}.}  \setlength{\tabcolsep}{0.85ex}
\begin{flushleft}
\begin{tabular}{cccccc}
\hline\noalign{\smallskip}
HJD\,245  & V       & EW    & HJD\,245  & V       & EW \\
          & (\kms)  & (\AA) &           & (\kms)  & (\AA) \\
\hline\noalign{\smallskip}
3202.6780 & $   4.5\pm3.9$ & 10.8 & 3202.8808 & $157.4\pm6.7$ &  4.8  \\
3202.6857 & $ -36.2\pm3.4$ &  9.1 & 3202.8884 & $161.1\pm6.2$ &  5.6  \\
3202.6954 & $ -64.0\pm3.6$ &  9.7 & 3202.8954 & $161.8\pm6.4$ &  6.2  \\
3202.7037 & $-103.4\pm3.5$ &  8.4 & 3202.9058 & $140.2\pm5.5$ &  7.0  \\
3202.7134 & $-128.7\pm4.1$ &  8.2 & 3202.9134 & $115.5\pm5.3$ &  8.3  \\
3202.7211 & $-150.9\pm5.1$ &  8.2 & 3202.9204 & $ 85.9\pm4.8$ &  8.1  \\
3202.7287 & $-204.6\pm6.8$ &  7.7 & 3202.9294 & $ 58.8\pm4.6$ &  7.9  \\
3202.7405 & $-205.2\pm6.1$ &  5.0 & 3202.9371 & $ 28.6\pm4.5$ & 10.0  \\
3202.7482 & $-218.7\pm6.1$ &  4.4 & 3202.9447 & $-31.9\pm6.4$ &  6.8  \\
3202.7836 &     -   & -0.6 & 3202.9537 & $ -44.8\pm4.7$ &  7.9  \\
3202.7912 &     -   & -2.3 & 3202.9614 & $ -82.5\pm5.0$ &  9.6  \\
3202.8013 &     -   & -3.0 & 3205.6857 & $-224.3\pm8.1$ &  1.5  \\
3202.8114 &     -   & -3.3 & 3205.7468 &     -   & -1.9  \\
3202.8197 &     -   & -2.9 & 3205.8114 & $217.3\pm10.4$ &  1.4  \\
3202.8273 &     -   & -3.0 & 3205.8760 & $28.6\pm4.7$   &  9.5  \\
3202.8405 &     -   & -0.2 & 3919.3887 & $-57.8\pm2.8$  &  8.7  \\
3202.8482 &     -   & -1.0 & 3920.3912 & $155.6\pm3.5$  &  4.6  \\
3202.8572 &     -   &  0.1 & 3921.3888 &     -   & -2.7  \\
3202.8648 & $153.1\pm10.3$ &  1.2 & 3922.3880 & $-226.1\pm4.7$ & 2.1 \\ 
3202.8725 & $140.8\pm8.7$ &  2.5 &              &         &       \\
\noalign{\smallskip}\hline
\end{tabular}
\end{flushleft}
Notes: The statistical error of the \Hb\ equivalent widths is
$\sim0.01-0.05$\,\AA, which is negligible. The systematic error,
however, is of the order $\sim1$\,\AA, depending on the details of how
the continuum flux is determined.
\end{table}

\subsection{Photometric and spectroscopic phase relations.}\label{s-phaserelations}
The assumption that the emission lines in HS\,1857+5144 originate on
the inner hemisphere of the secondary star as a result of strong
irradiation from the hot primary star makes specific predictions on
the relative phases of the photometric and spectroscopic variability.
At superior conjunction of the secondary star when the irradiated side
faces the observer, $\varphi=0.5$, the system appears brightest; the
radial velocity of the secondary star is zero and crossing from
red-shifted to blue-shifted velocities; the emission-line strength is
at the maximum, and vice versa for the inferior conjunction of the
secondary star at phase zero. Hence, one would expect an agreement in
phase between the light curve and the equivalent width variation, and
a 0.25~phase shift between those two parameters and the radial
velocity curve \citep[e.g.][]{thorstensenetal78-1,
thorstensenetal94-1, thorstensenetal96-2, vennes+thorstensen96-1,
oroszetal99-1, hillwigetal00-1, kawkaetal02-1}.
Figure\,\ref{f-fold_hs1857} shows the radial velocity variation of the
average of the Balmer lines and the equivalent widths of \Hb\ as well
as the $B$- and $R$-band light curves folded over
Eq.\,\ref{e-ephemeris}.  The phase offset between the $B$- and
$R$-band light curves with respect to the equivalent width variation,
as determined from sine fits, is $\sim0.004$ for the $R$-band and
$\sim0.018$ for the $B$-band. The larger phase offset of the $B$-band
light curve is probably related to the fact that it does not cover the
entire orbital cycle, and hence the sine fit results in larger
uncertainties. The phase of the folded equivalent width variation lags
that of the radial velocity curve by $0.25\pm0.01$ orbital cycle.

The phase-dependent behaviour of the emission lines, and the relative
phases of the photometric, radial velocity, and equivalent width variations
found in our data corroborate the hypothesis of the emission lines
in HS\,1857+5144 originating on the inner face of the secondary
star illuminated by the hot white dwarf.

\section{Stellar components}\label{s-stellar_components}
\subsection{Light curve solution}\label{s-lc_solution}
In order to determine additional constraints on the system parameters
from the observed reflection effect we modelled the light curves of
HS\,1857+5144 with the 'PHysics Of Eclipsing BinariEs' program
\texttt{PHOEBE}\footnote{http://phoebe.fiz.uni-lj.si/}
\citep{prsa+zwitter05-1}, which is built on top of the widely used WD
code \citep{wilson+devinney71-1, wilson79-1, wilson90-1}. We
simultaneously fitted the $R$- and $B$-band data obtained at Kryoneri
Observatory under the following assumptions: (a) circular orbits and
synchronous rotation of the secondary star; (b) stellar surface
temperature and brightness were computed assuming blackbody emission; (c)
a detailed calculation of the reflection effect was adopted; (d)
linear limb darkening was chosen, where the limb darkening
coefficient was interpolated from \citet{claret00-1}; (e) gravity
darkening exponents of 1 \citep{vonzeipel24-1} and 0.32
\citep{lucy67-1} were used for radiative and convective stars,
respectively; (f) no contribution fluxes from a spot or third light
were applied.

In our analysis we tested a wide range of white dwarf mass, covering
$\Mwd=0.3-1.4$\,\Msun.  We then assumed an M-type companion star,
testing the whole range of spectral type M9--M0\,V, corresponding to
masses, radii, and temperatures of $\Msec\simeq0.07-0.53$\,\Msun,
$\Rsec\simeq0.11-0.56$\,\Rsun, and $\Tsec\simeq2300-3800$\,K,
interpolated from Rebassa-Mansergas et al. (in preparation). Earlier
spectral types than M0\,V would imply extremely massive white dwarfs
(Sect.\,\ref{s-kcorrection}) which are excluded by the spectral fit
(Sect.\,\ref{s-spectralfit})\footnote{A strict upper limit on the mass
of the companion comes from the fact that it is not Roche-lobe
filling. Using $\bar{\rho}\simeq107P_\mathrm{orb}^{-2}(\mathrm{h})$
\citep[e.g.][]{eggleton83-1}, with $\bar{\rho}$, the average density of
the donor, and $P_\mathrm{orb}(\mathrm{h})$, the orbital period in hours
for a Roche-lobe filling star, the maximum mass of a main-sequence
companion in HS\,1857+5144 is $\Msec<0.72$\,\Msun, corresponding to a
spectral type K3\,V or later.}.  This approach allows us to search for possible
solutions over a large range of possible mass ratios,
$q=\Msec/\Mwd\simeq0.05-1.77$. For each input $q$ we fixed
$\Porb=0.26633$\,d and \Tsec\ according to the selected spectral type
of the companion star. The following parameters were free in the fits:
$q$, white dwarf temperature (\Twd), orbital inclination ($i$),
surface potentials for both components, and albedo of the secondary
star ($ALB2$). The fits for early type companions, M2--M0\,V, do not
reproduce well the observed $B$ and $R$ light curves, which supports
the exclusion of early-type donors outlined above.  We found that,
independently from the details of a given fit, the system must contain
a hot white dwarf with $\Twd>60\,000$\,K and a cool component with an
albedo higher than that of a normal M star ($ALB2>0.5$) to reproduce
the large amplitude observed in the light curves. Such high \Twd\ is
also confirmed by the spectral fit to the WHT faint-phase spectrum in
the following section. Fitting the $B$- and $R$-band light curves
alone provides a fairly large range of possible system parameters,
$\Mwd\simeq0.3-1.4$\,\Msun, $\Msec\simeq0.066-0.367$\,\Msun, (spectral
type M9--M3\,V), $i\sim40\degr-60\degr$, and
$\Twd\sim60\,000-100\,000$\,K. For a given $\Msec$, a more massive
white dwarf requires a larger inclination and a higher $\Twd$.

Figure\,\ref{f-fold_hs1857} (bottom two panels) presents the
corresponding $R$- and $B$-band synthetic light curves from the program
\texttt{PHOEBE} for $\Mwd=0.72$\,\Msun, $\Msec=0.21$\,\Msun,
$i=53\degr$, and $\Twd=100\,000$\,K, along with the observed light
curves folded over the ephemeris in Eq.\,\ref{e-ephemeris}. The choice
of this particular set of parameter is detailed below in
Sect.\,\ref{s-combinedconstraints}, but other fits in the parameter range
given above fit the data equally well.

\subsection{Spectral fit}\label{s-spectralfit}
We performed a spectral fit to the faint-phase WHT spectrum of
HS\,1857+5144 to obtain an independent estimate of \Mwd\ and \Twd,
using both a grid of LTE pure-hydrogen models \citep{koesteretal05-1},
and a grid of NLTE models with a variety of He
abundances\footnote{http://astro.uni-tuebingen.de/$\sim$rauch}.  We
fitted the \Hb\ and \Hg\ absorption lines after normalising the
continua of the observed data and the model spectra in the same way
using a third-order polynomial. The fits suggest
$80\,000\,\mathrm{K}\la\Twd\la 100\,000$\,K and $7.5\la\log\,g\la 8.5$
(corresponding to $0.6\,\Msun\la\Mwd\la 1.0$\,\Msun) for the LTE models, and
$70\,000\,\mathrm{K}\la\Twd\la 100\,000$\,K and $8.0\la\log\,g\la 8.5$
(corresponding to $0.7\,\Msun\la\Mwd\la 1.0$\,\Msun) for the NLTE models.
Figure\,\ref{f-balmerfit} shows the best LTE fit to \Hb\ and \Hg\ for
$\Twd=100\,000$\,K and $\log\,g=8.0$.  These numbers should be
considered as rough estimates only, as the optical spectrum of
HS\,1857+5144 is contaminated by flux from the companion star, which
is very difficult to quantify. A more reliable temperature and mass
estimate would require far-ultraviolet data, where the white dwarf
dominates the emission of the system \citep{goodetal04-1}. We conclude
from this qualitative spectral analysis that the white dwarf in
HS\,1857+5144 is indeed very hot and, given the high surface gravity
preferred by the fits, is more likely to be a white dwarf than a
subdwarf. The detection of \Line{He}{II}{4686} then qualifies the
primary as a DAO white dwarf.

\begin{figure}
\centerline{\includegraphics[angle=-90,width=\columnwidth]{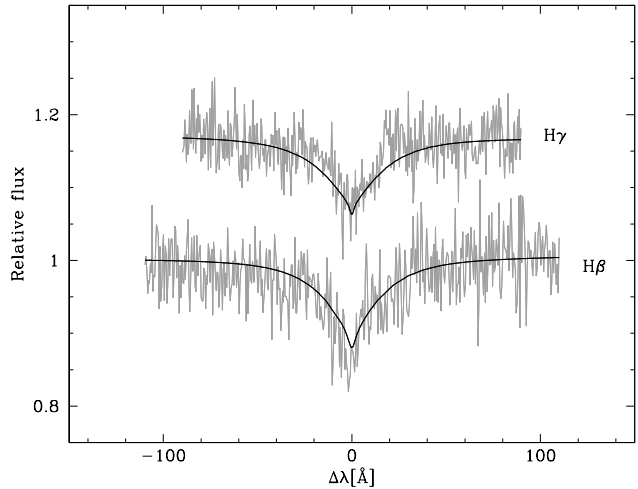}}
\caption{\label{f-balmerfit} \Hb\ and \Hg\ from the WHT faint-phase
spectrum ($\varphi=0.02$) fitted with an LTE model for $\Twd=100\,000$\,K
and $\log\,g=8.0$.}
\end{figure}

\subsection{$K$-correction and mass ratio-inclination constraints\label{s-kcorrection}}
As mentioned already in Sect.\,\ref{s-rv}, the emission lines in
HS\,1857+5144 trace the orbit of the centre of light of the
illuminated hemisphere of the secondary star, and not its centre of
mass. Hence, for a dynamic assessment of the binary parameters, the
measured velocity \Kem\ (Sect.\,\ref{s-rv}) has to be corrected
accordingly. The radial velocity amplitude of the secondary star's
centre of mass, \Kseccor\, can be expressed according to
\citet{wade+horne88-1} as
\[\Kseccor=\frac{\Kem}{1-(1+q)({\Delta}R/a)}~~,\]
where $\Delta R$ is the displacement of the centre of light from the
centre of mass of the secondary star, with $0 \la \Delta R \la \Rsec$
($\Delta R=0$ implies that the centre of light coincides with the
centre of mass of the secondary star, whereas $\Delta R=\Rsec$ is the
maximum possible displacement, where all the light comes from a small
region on the secondary star closest to the primary star).  Assuming
that the emission due to irradiation is distributed uniformly over the
inner hemisphere of the secondary star, and zero on its unirradiated
face, $\Delta R=(4/3\pi)\Rsec$ \citep{wade+horne88-1, woodetal95-3,
oroszetal99-1, vennesetal99-2}.

The expected radial velocity of the secondary star (\Kseccal) is 
\[\Kseccal=\frac{2\pi a \sin i}{\Porb(1+q)}~~,\]
where $a$ is the binary separation. Equating $\Kseccor=\Kseccal$ then
gives a unique $q$ for a given choice of $i$, Hence, a fixed value of
$i$ projects onto a one-dimensional curve within the ($\Mwd,\Msec$)
plane, and for the possible range of parameters considered here, those
curves are nearly straight lines.

\subsection{\label{s-combinedconstraints}Combined constraints}
In Sect.\,\ref{s-lc_solution}-\ref{s-kcorrection} above we have
outlined what type of constraints on the system parameters of
HS\,1857+5144 can be derived from the observed light curves, radial
velocity variations, and the spectrum of the primary star. Here, we will
combine all those independent constraints. 

In a first step, we impose a range of $i=40^\circ-60^\circ$, as
suggested by the set of \texttt{PHOEBE} fits to the $B$- and $R$-band
light curves, on the combinations of $(\Mwd,\Msec)$ which are
consistent with the corrected radial velocity \Kseccor\ of the
secondary star. The resulting parameter range is indicated by the gray
shaded area in Fig.\,\ref{f-qconst}. In a second step, we inspected
the individual light curve fits from the grid of \texttt{PHOEBE} runs,
and require the inclination of a model for a given $(\Mwd,\Msec)$
to fall within $\pm5^{\circ}$ of the corresponding inclination
constraint from the radial velocity of the secondary star. We
introduce this ``fuzziness'' in inclination as a measure to account
for systematic uncertainties within the $K$-correction and the light
curve fits. Possible combinations of $(\Mwd, \Msec)$ are indicated by
filled circles in Fig.\,\ref{f-qconst}, and trace a somewhat narrower
band than the initial $i=40^\circ-60^\circ$ constraint. A final
constraint comes from the spectral fit of the WHT faint phase spectrum
(Sect.\,\ref{s-spectralfit}), which implied $0.6\,\Msun \la \Mwd \la
1.0\,\Msun$, shown as vertical dashed lines in Fig.\,\ref{f-qconst}. 

\begin{figure}
\centerline{\includegraphics[angle=-90,width=\columnwidth]{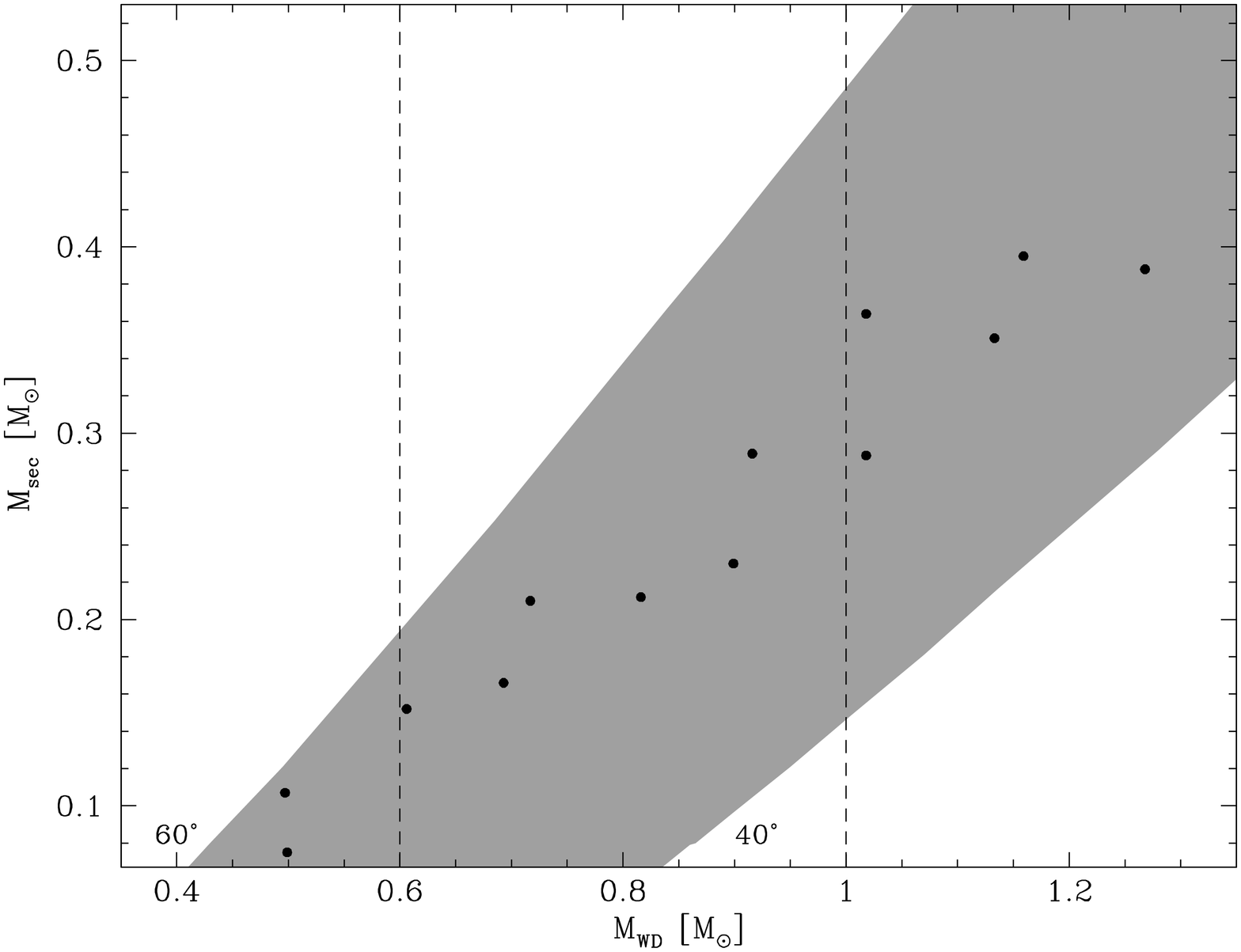}}
\caption{\label{f-qconst} Photometric and spectroscopic constraints on
(\Mwd, \Msec). Gray shade area represents possible dynamical solutions
from the 'K-corrected' radial velocity of the secondary star for
$\Kem=185$\,\kms\ and for $i=40\degr-60\degr$. Filled circles represent
possible solutions from the light curve analysis for given (\Mwd, \Msec)
in which $i$ agrees with spectroscopic constraint within $\pm5\degr$. Dashed
lines are upper and lower limits on \Mwd\ derived from the spectral fit
(see Sect.\,\ref{s-combinedconstraints} for details).}
\end{figure}

The combination of all constraints suggests $0.15\,\Msun \la\Msec
\la0.30$\,\Msun\ (spectral type M6--M4\,V), $0.6\,\Msun \la \Mwd \la
1.0\,\Msun$, $70\,000\,\mathrm{K} \la\Twd \la100\,000$\,K, and
$45\degr \la i \la 55\degr$. A substantial improvement on this set of
parameters will require measuring the radial velocity amplitude of the
white dwarf, \Kwd, and determining an accurate temperature for the
primary. Both types of measurements could be easily obtained from
time-resolved ultraviolet spectroscopy.

\begin{figure}
\centerline{\includegraphics[angle=-90,width=\columnwidth]{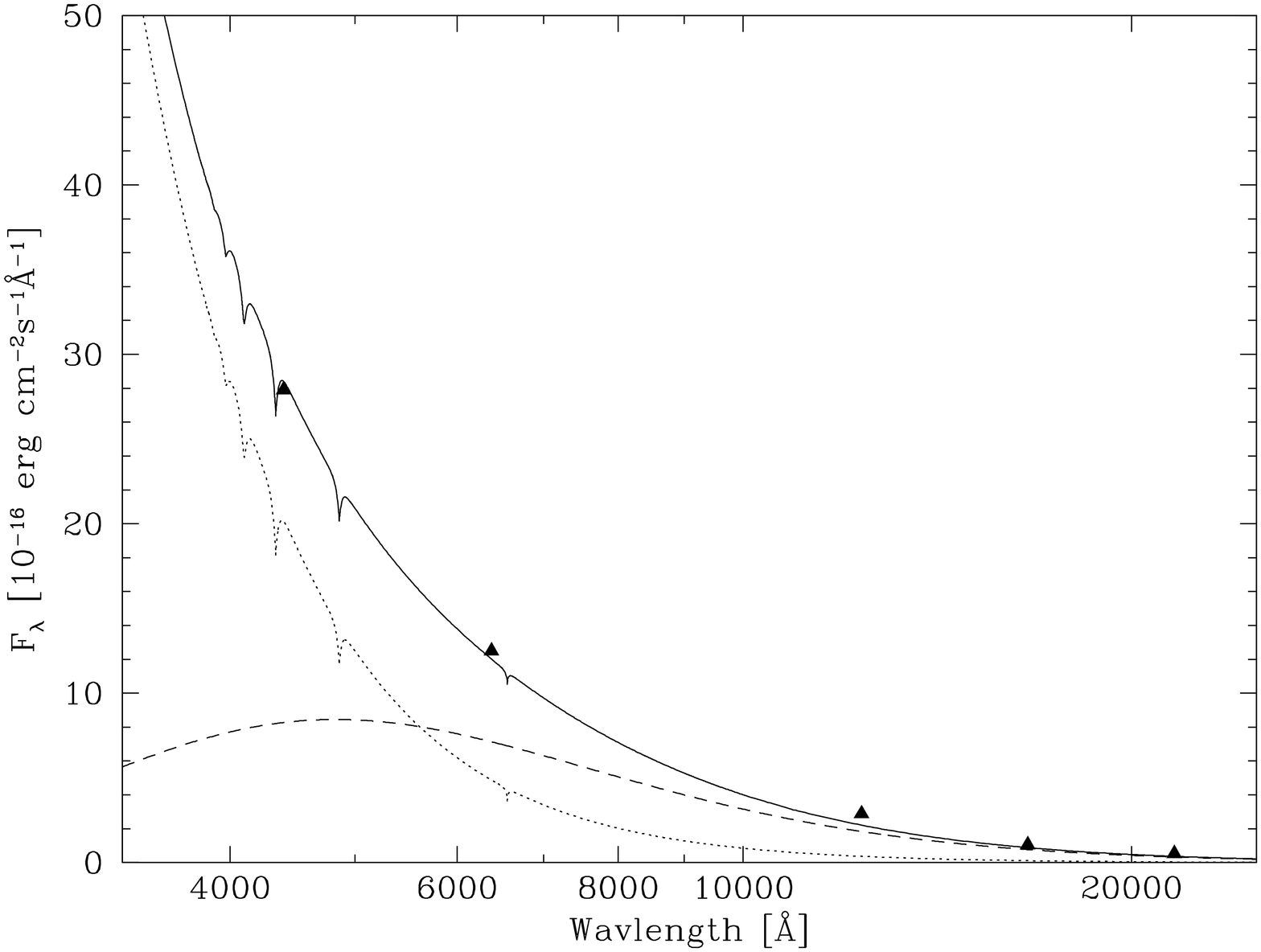}}
\caption{\label{f-sed} The $BRJHK_\mathrm{s}$ fluxes of
HS\,1857+5144 (filled triangles) at orbital maximum ($\varphi\simeq0.5$)
along with an example of a plausible fit (solid line) with the sum of
a white dwarf spectrum (dots) and a blackbody (dash line) representing
the heated side of the low-mass companion, assuming $\Twd=70\,000$\,K,
$\Rwd=1.3\times10^{9}$\,cm, $\Rsec=1.3\times10^{10}$\,cm, and
$\Tsec=6000$\,K, at a distance of 460\,pc (see Sect.\,\ref{s-2mass}
for details).}
\end{figure}

\subsection{2MASS magnitudes}\label{s-2mass} 
HS\,1857+5144 is detected in the 2MASS \citep{skrutskieetal06-1} at
$J=15.09\pm0.04$, $H15.05\pm0.08$, $K_\mathrm{s}=14.76\pm0.14$. Our
ephemeris is good enough to establish the orbital phase of the 2MASS
data, which is very close to orbital maximum, $\varphi\simeq0.5$. In
order to test to what extent the optical-infrared spectral energy
distribution (SED) of HS\,1857+5144 is compatible with the stellar
photometry and spectroscopy, we modelled the $BRJHK_\mathrm{s}$
parameter range obtained from the analysis of the time-resolved
magnitudes with the sum of a white dwarf spectrum from
\citep{koesteretal05-1} and a blackbody representing the contribution
of a (heated) low-mass companion. The data leaves some freedom in the
exact parameters, however, as an example, a fairly good fit is
achieved for $\Twd=70\,000$\,K, $\Rwd=1.3\times10^{9}$\,cm,
corresponding to $\Mwd\simeq0.6$\,\Msun, $\Rsec=1.3\times10^{10}$\,cm,
corresponding to spectral type of M6, and $\Tsec=6000$\,K, at a
distance of 460\,pc (Fig.\,\ref{f-sed}). This solution coincides well
with the photometric and spectroscopic constraints in
Fig.\,\ref{f-qconst}. Overall our simple model confirms the stellar
parameters established in the previous sections, with a slight
preference for a white dwarf mass in the range $\Mwd=0.6-0.8$\,\Msun,
and a companion star with a radius near the lower end of the
determined range, i.e. corresponding to a spectral type M6--M5, and a
distance to the system of 290--460\,pc. 

\begin{table*}[t]
\caption[]{PCEBs with a large reflection effect.\label{t-reflection}}
\setlength{\tabcolsep}{0.9ex}
\begin{flushleft}
\begin{tabular}{lcccccccccc}
\hline\noalign{\smallskip}
Object & \Porb & SP1 &  SPsec & \Tone & \Mone & \Msec &     
\multicolumn{3}{c}{Reflection effect [mag]} & Ref. \\ 
&  [d]  &  &  & [K] & [\Msun] & [\Msun] & $B$ & $V$ & $R$ & \\
\hline\noalign{\smallskip}
NN\,Ser & 0.130 & DA & M4.75\,V & $57\,000\pm3000$ & $0.54\pm0.05$ & $0.150\pm0.008$ & 0.33 & 0.49 & 0.772 & 1, 2 \\
HS\,1857+5144 & 0.266 & DAO & $\sim$M6-M4\,V & $\sim70\,000-100\,000$ & $\sim0.6-1.0$ & $\sim0.15-0.30$ & 0.7  & & 1.1 & 3 \\
TW\,Crv & 0.328 & sdO & M\,V & $105\,000\pm20\,000$ & $\sim0.55-0.61$ & $<0.3$ & 0.74 & 0.85 & 0.93 & 4, 5 \\
KV\,Vel & 0.357 & sdO,PN & M\,V & $77\,000\pm3000$ & $0.63\pm0.03$ & $0.23\pm0.01$ & 0.49 & 0.55 & 0.61 & 6 \\
V477\,Lyr & 0.472 & sdOB,PN &	        & $60\,000\pm10\,000$ & $0.51\pm0.07$ & $0.15\pm0.02$ & 0.5 & 0.6 & & 7, 8\\
V664\,Cas & 0.582 & sdO,PN & K5-F5\,V & $83\,000\pm6000$ & & & & 1.15 & & 9, 5\\
VW\,Pyx   & 0.676 & sdO,PN & & $85\,000\pm6000$ & & & & 1.36 & & 10, 11\\
Abell\,65 & 1	 & sd?,PN     &	        & $\sim80\,000$ & & & & $>0.5$ & & 12, 13 \\
BE\,UMa & 2.291 & sdO/DAO,PN & K4-3\,V & $105\,000\pm5000$ & $0.70\pm0.07$ & $0.36\pm0.07$ & & $\sim1.3$ & & 14, 15, 16\\
\noalign{\smallskip}\hline
\end{tabular}
\end{flushleft}
References: (1) \citet{haefner89-1}; (2) \citet{haefneretal04-1}; (3)
this work; (4) \citet{chenetal95-1}; (5) \citet{exteretal05-1}; (6)
\citet{hilditchetal96-1}; (7) bond+grauer-87; (8)
\citet{pollacco+bell94-1}; (9) shimanskiietal04-1; (10)
\citet{kohoutex+schnur82-1}; (11) \citet{exteretal03-1}; (12)
\citet{bond+livio90-1}; (13) \citet{walsh+walton96-1}; (14)
\citet{fergusonetal87-1}; (15) \citet{woodetal95-3}; (16)
\citet{fergusonetal99-1}
\end{table*}


\section{Discussion}\label{s-discussion} 
The analysis presented in Sect.\,\ref{s-spectralfit} suggests that
HS\,1857+5144 contains a hot white dwarf with
$\Twd\simeq70\,000-100\,000$\,K. The implied cooling age of the white
dwarf is $1.2-6\times10^{5}$\,yr (\citealt{bergeronetal95-2}, and
Bergeron 2002, private communication), making HS\,1857+5144 one of the
youngest PCEBs known so far. Following the prescription of
\citet{schreiber+gaensicke03-1}, and assuming the range of system
parameters established in Sect.\,\ref{s-stellar_components} as well as
``classical'' magnetic braking for the angular momentum loss mechanism,
we estimate the period at which HS\,1857+5144 left the common envelope
phase to be $P_\mathrm{CE}\simeq0.266334-0.266345$\,d, very close to
its present orbital period. HS\,1857+5144 will evolve within the next
$\sim0.4-1.3\times10^{10}$\,yr into a semidetached CV configuration,
and start mass transfer at an orbital period of $\simeq0.08-0.13$\,d,
i.e. within or below the period gap. The large uncertainties on the
future evolution are a consequence of the limited constraints on the
system parameters. Additional systematic uncertainties in the actual
strength of angular momentum loss from the orbit have not been taken
into account.

Among $\sim40$ previously known PCEBs listed in
\citet{schreiber+gaensicke03-1}, and \citet{shimanskyetal06-1},
only 8 systems display a reflection effect comparable to that of
HS\,1857+5144 (Table\,\ref{t-reflection}). All those systems contain
extremely hot white dwarfs or subdwarfs, and all are very young PCEBs
that may serve as observational probes into our understanding of
common envelope evolution. 

A large reflection effect is expected for those PCEBs containing a hot
subdwarf, because of the larger luminosity compared to a white dwarf
of the same temperature, and indeed the majority of known PCEBs with a
large reflection effect have sdO primary stars.  So far only one PCEB,
HS\,1136+6646, containing a hot white dwarf similar to HS\,1857+5144
\citep{singetal04-1} is known. The secondary star in HS\,1136+6646 has
been suggested to be a K7--4\,V star on the basis of its spectral type,
which appears too early for the estimated mass of
0.34\,\Msun. However, \citet{singetal04-1} discuss the possibility
that the secondary is overluminous as it is still out of thermal
equilibrium after accreting at a high rate during the common envelope
phase. The amplitude of the reflection effect in HS\,1136+6646 is much
lower than in HS\,1857+5144, which is consistent
with its longer orbital period of 0.84\,d. The other system most
similar to HS\,1857+5411 is BE\,UMa, which has been classified as a
borderline object between an sdO subdwarf and a DAO white dwarf
\citep{liebertetal95-1, fergusonetal99-1}, and is associated with a
planetary nebula. It is interesting to note that six out of the nine
systems listed in Table\,\ref{t-reflection} are still embedded in a
planetary nebula. Our long-slit spectroscopy of HS\,1857+5144 does not
reveal any noticeable trace of extended emission around \Ha, though a
deep \Ha\ image testing for remnant nebular emission would be
useful. Similarly, no sign of extended \Ha\ emission around HS\,1136+6646
has been observed. While the majority of very young PCEBs are still
embedded in their planetary nebulae/common envelopes, the discovery of
HS\,1857+5144 and HS\,1136+6646 suggests that the envelope can be
dispersed rather quickly.

\section{Conclusions}
We have identified a pre-CV, HS\,1857+5144, containing a hot DAO white
dwarf and a low mass M\,V star with an orbital period of
$\Porb=383.52$\,min. The light curves of HS\,1857+5144 display a very
large reflection effect with peak-to-peak amplitude of 0.7 and 1.1 mag in the
$B$ and $R$ bands, respectively. Combining the results of our
spectroscopic and photometric analysis, we constrain the system
parameter to $0.15\,\Msun \la\Msec \la0.30$\,\Msun\ (spectral type
M6--M4\,V), $0.6\,\Msun \la \Mwd \la 1.0\,\Msun$,
$70\,000\,\mathrm{K}\la\Twd \la100\,000$\,K, and $45\degr \la i \la
55\degr$. Taking these parameters at face value HS\,1857+5144 is one
of the youngest PCEBs known so far and has just emerged from the post common
envelope phase. The system will eventually evolve into a cataclysmic
variable, and start mass transfer within or below the $2-3$\,h orbital
period gap.

\acknowledgements AA thanks the Royal Thai Government for a
studentship. BTG and PRG were supported by a PPARC Advanced Fellowship
and a PDRA grant, respectively.  The HQS was supported by the Deutsche
Forschungsgemeinschaft through grants Re\,353/11 and Re\,353/22. We
thank John Southworth for reducing the WHT spectra. Tom Marsh is
acknowledged for developing and sharing his reduction and analysis
package \texttt{MOLLY}. We thank the anonymous referee for
his/her comments that lead to an improved presentation of the
paper. This publication makes use of data products from the Two Micron
All Sky Survey, which is a joint project of the University of
Massachusetts and the Infrared Processing and Analysis
Center/California Institute of Technology, funded by the National
Aeronautics and Space Administration and the National Science
Foundation.

Based in part on observations collected at the Centro Astron\'omico
Hispano Alem\'an (CAHA) at Calar Alto, operated jointly by the
Max-Planck Institut f\"ur Astronomie and the Instituto de
Astrof{\'\i}sica de Andaluc{\'\i}a (CSIC); on observations made with
the 2.7m telescope at the McDonald Observatory of the University of
Texas at Austin (Texas); on observations made at the 1.2m telescope,
located at Kryoneri Korinthias, and owned by the National Observatory
of Athens, Greece; on observations made with the William Herschel
Telescope, operated on the island of La Palma by the Instituto de
Astrof{\'\i}sica de Canarias (IAC) at the Spanish Observatorio del
Roque de los Muchachos; on observations made with the IAC80 telescope,
operated on the island of Tenerife by the IAC at the Spanish
Observatorio del Teide of the IAC; on observations made with the
Optical Groud Station telescope, operated on the island of Tenerife by
the European Space Agency, in the Spanish Observatorio del Teide of
the IAC.

\bibliographystyle{aa}

\end{document}